# Recompositing of Vast Irregularly - Sampled Seismic Data via Compressed Sensing Framework: An FPOCS Based on Seislet Transform Approach

Hussein Muhammed[iD]1,2,3*

1. Key Laboratory of Deep Oil & gas, China University of Petroleum (East China), Qingdao, 266580, China;
2. Shandong Provincial Key Laboratory of Reservoir Geology, China University of Petroleum (East China), Qingdao, 266580, China;
3. Centre for Seismological Phenomena, Department of Geology, Faculty of Science, University of Khartoum, Khartoum 11115, Sudan.

*Corresponding author's address: Dept., of Geophysics, School of Geosciences, China University of Petroleum (East China), No. 66, West Changjiang Rd., Huangdao District, Qingdao 266580, Shandong, P.R China.

(Email:mhmd_hussein@s.upc.edu.cn).

**Abstract**

Acquiring seismic data from irregular topographic surface is oftently oppressed by irregular and nonequivalent source-receiver arrays and even more it yields bad traces after storing the original signal. In the light of preprocessing seismic data, we have to extract out most of the given signal, thus further processing and interpretation can obtain extremely accurate outcomes. We applied Compressed Sensing theorem on Sigmoid vast irregularly-sampled seismic data based on the fast projection onto convex sets (FPOCS) algorithm with sparsity constraint in the seislet transform domain, which gives faster convergence than other conventional methods and is preserving an optimum signals recovery. The FPOCS seislet transform approach can achieve accurate and high data recovery results than other methods because of a much sparser structure in the seislet transform domain as demonstrated. Moreover, FPOCS algorithm is also efficient in minimizing the number of required iterations to achieve optimum data refilling.

**Introduction**

Compressed Sensing aims to reconstruct a full high-resolution image, or whatever objective form of signals, from a dramatic subsampling of the given data by assuming there is a universal picture or signal matrix that contains out input data. Even if we only have a small number of random pixels from that image we are able to infer the active Fourier coefficients in that image. The principal framework of it encompasses solving a least-squares minimization riddle with an $L_1$ norm condition of the recomposed model, which requires compromising a least-square data-misfit constraint and a sparsity constraint over the reconstructed model. There are two conditions to achieve Compressive Sensing which are; we have to measure seismic data and eliminate about half of it to get fewer signals in the Fourier basis and the second one; these seismic measurements must be collected randomly to allow signals recovering.

There are several methods can be composed to solve the problem such as: iterative shrinkage thresholding (IST) and the projection onto convex sets (POCS) which are common approaches used to solve the minimization problem in the seismic data processing field. The key step in finding the sparest solution to the optimization problem lies in the technique of converting to Fourier transform to find the sparsest matrix or solve the combinatorial hard problem to satisfies the system of equation by an infinite trials of choices. The actual development of this technology started in mid 2000s (after Candès et al. 2006; Donoho, 2001; 2006; Ying & Hao, 2010; Donoho & Tanner, 2010; Herrmann, 2010). The FPOCS algorithm (Gan et al. 2016) is equivalent to the fast iterative shrinkage-thresholding algorithm (FISTA) by Beck and Teboulle (2009). The seislet transform is sparser than other auxiliary sparse transforms (after Chen et al., 2014; Fomel and Liu, 2010).

Baraniuk & Steeghs (2017) addressed various topic and discussed several applications to seismic data acquisition and processing while Elzanaty et al. (2018) gave an analysis of the restricted isometry constant (RIC) of finite dimensional Gaussian measurement matrices to impart a tight lower bound on the maximum sparsity order of the given signal which in turn allows signal recovery with an already set

# FPOCS Based on Seislet Transform Compressed Sensing Algorithm

target probability. A compressive sensing based method to regularize non-stationary VSP data which obtains improved data reconstructions, is proposed by Yu et al. (2020). This paper aims at introducing Compressed Sensing method and a fast calculation method for restoring seismic data. The relevant concepts and theories are presented and a synthetic example is given to demonstrate the FPOCS algorithm in comparison with well know algorithms.

## Compressed Sensing Framework

The following equations are generalized formulation of our framework.

Any compressible image/signal $x \in R^n$ may be written as a sparse vector $s \in R^n$ in a $\Psi$ Fourier transform basis as:

$$x = \Psi.s \quad (1)$$

Assume that we have some seismic measurements and $C$ is measurement matrix represents a set of linear measurements, $y$ which is a function of the compressible image/signal $x$, then:

$$y = C x$$
$$y = C\Psi.s \quad (2)$$

Equation (2) is an optimization or inverse problem due the reversibility of the term $\Psi$ (Fourier transform and its inverse) thus we solve for the sparsest solution that satisfies the whole system of equations by finding $s$.

## FPOCS Algorithm

The FPOCS approach (Gan et al. 2016) is trying to solve the following system of equations:

$$\min_{d} \left\| d_{obs.} - S d \right\|_2^2 + \lambda \left\| A d \right\|_1 \quad (3)$$

where $d_{obs.}$ is the collected seismic data, $S$ is the sampling operator, $d$ is the unknown estimated seismic data and $A$ is the sparsity-promoting transform.

The projection onto convex sets conventional algorithm is globally used one for recompositioning incomplete seismic traces, especially in the case of vast irregularly-sampled seismic data casted onto conventional grids. Liang et al. (2014) stated that; the analysis-based approach underlines the sparsity of the canonical transformed coefficients; thus it is likely to restore seismic data with smooth regions; while the synthesis-based method realizes the sparsest approximation of the given seismic data in the transformed domain.

## Seismic Data Examples

In order to demonstrate the idea of Compressed Sensing (CS) we use Sigmoid reflectivity model (Figure1) and decimated the traces to illustrate how CS efficient can be in reconstructing vast irregularly-sampled. Figure 2(a) is a reconstructed output reflectivity image after applying projection onto convex sets (POCS) with $f-k$ thresholding algorithm while 2(b) shows a restored image after implementing Fast-POCS with $f-k$ thresholding. We see both algorithms have restored valuable parts of the initial reflectivity model. On the other hand, applying the seislet transform can increase the convergence of the whole framework and decrease the number of iterations to match the data as seen in Figure 3(a) since we apply POCS with seislet thresholding. In addition to that, the algorithm that runs very fast was the Fast-POCS (FPOCS) with seislet thresholding Figure 3(b). In general, all algorithm works significantly without any issues, however the last one is the fastest and it preferred in industry.

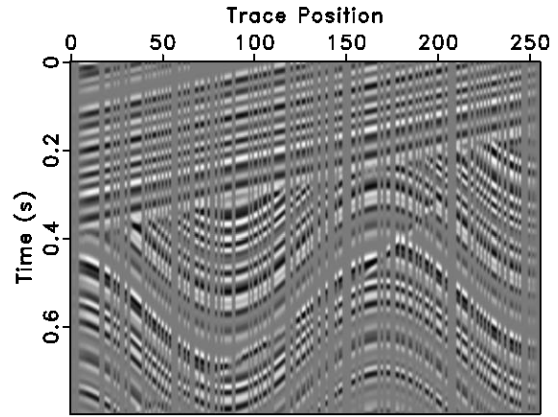

Figure1: Sigmoid synthetic reflectivity model (after Claerbout, 2014) with about 45% missing traces.

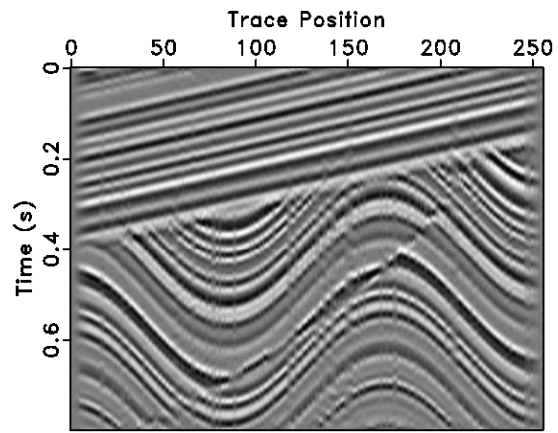

(a)

# FPOCS Based on Seislet Transform Compressed Sensing Algorithm

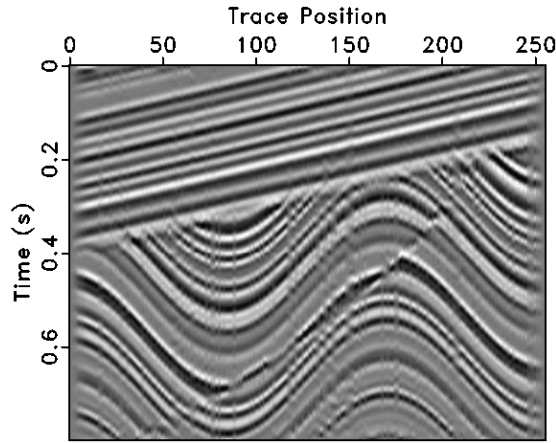

(b)

Figure2: CS through: (a) POCS, (b) $f-k$ thresholding algorithm.

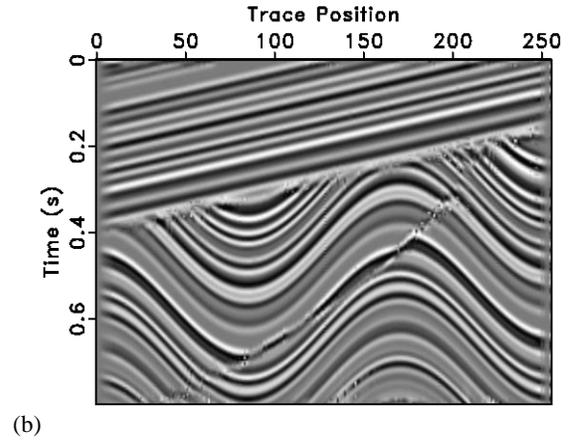

(b)

Figure3: CS through: (a) POCS with seislet thresholding, (b) FPOCS with seislet thresholding.

## Conclusion

The FPOCS algorithm for applying Compressed Sensing on seismic data via sparsity constraint in seislet transform domain will have huge potentialities when implementing it on field data. The FPOCS can obtain much faster convergence than conventional POCS, which can potentially make the seislet-based POCS approach applicable in practice according to the efficiency acceleration. This conclusion can guide us to use different iterative approach according to the noise level in the data. The CS based on seislet transform can obtain optimum data recovery results than $f-k$ transform based algorithms because of a much sparser structure in the seislet transform domain. We have used synthetic data examples to demonstrate the advantages of using CS seislet-based FPOCS approach.

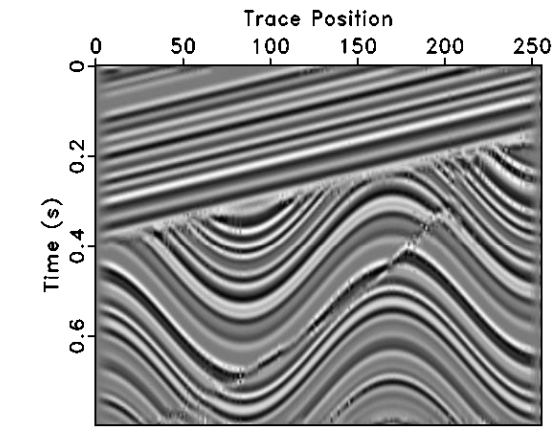

(a)

## Keywords

Compressed Sensing, Vast Irregularly-sampled seismic data, Fourier Transform, Seislet transform.

## Acknowledgment

The research is funded by: National Natural Science Foundation of China (41574098 & 41630964). We thank the colleagues and students within SWPI laboratory for their weekly discussion. Special thanks to the creators of Madagascar software and Prof. Dr. Jeffery Shragge (center for wave phenomena, CSM, CO, USA) for his valuable discussion and technical assistant.

# FPOCS Based on Seislet Transform Compressed Sensing Algorithm